# Physical Properties of Materials and Testing Methods

*Ana Arauzo*
Universidad de Zaragoza, Pedro Cerbuna 12, 50009 Zaragoza, Spain

**Abstract**
The use of materials in particle accelerator applications implies the consideration of physical properties in different environments than usual industrial applications. Cryogenic temperatures, high radiation, medium to high vacuum and high magnetic fields are requirements for which the behavior of materials has to be evaluated. In this lecture, Thermal, Electrical and Magnetic properties are briefly introduced, highlighting the mechanisms governing variation with temperature and magnetic field. Standards and testing methods for different properties are described, illustrated with practical examples. With regard to the characterization of thermal properties of a material, testing methods of heat capacity and thermal conductivity are described, with details of the main working principle and measurement examples. Determination of electrical resistivity and the residual resistivity ratio (RRR) are the main characteristics of electrical properties, together with specific characterization of superconducting materials. Testing magnetic properties deserve special attention; different methods are commonly employed to characterize non-magnetic and magnetic materials. In magnetic metals and ferrites non-linear behavior and hysteresis effects need to be considered in order to assess their suitability for applications or to model material performance.

## 1   Introduction

The use of materials in particle accelerator applications requires consideration of their physical properties in environments that differ significantly from typical industrial conditions. These materials must be characterized under extreme conditions, including high radiation doses that cause cumulative damage, cryogenic temperatures as low as 1.9 K (as in the superfluid helium used in superconducting magnets), steep temperature gradients, medium to high vacuum, and strong magnetic fields up to 8.3 tesla, approximately 200,000 times the Earth's magnetic field. Material behavior under such conditions must be thoroughly evaluated.

For example, in the LHC at CERN, the dipole magnets used to bend particle trajectories are among the most complex components of the accelerator. There are 1232 main dipoles, each 15 meters long and weighing about 35 tonnes, made from materials capable of withstanding high magnetic fields, mechanical loads, cryogenic temperatures, and thermal gradients, while meeting exceptionally strict requirements for magnetic field uniformity and positioning accuracy. Additionally, the LHC requires an extremely robust and unprecedented collimation system. Approximately 30% of the beam is lost in the cleaning insertions, making these areas among the most radioactive in the entire LHC ring. The system uses durable primary and secondary collimators made from carbon-fiber composite, along with tungsten absorbers to protect the superconducting magnets located downstream of the warm insertions.

The main physical properties that must be tested under extreme conditions, particularly temperatures below 10 K and magnetic fields above 2 T, include electrical properties (such as resistivity and electrical transport), thermal properties (including heat capacity and thermal conductivity), and magnetic properties (the material response to applied magnetic fields). Suppliers typically provide material specifications for standard industrial conditions, but accelerator applications require dedicated instrumentation for characterization under these specific extreme environments. These low-temperature and high-field conditions can be reproduced using scientific instruments developed for materials research. The measured physical properties can be used for performance modelling or to assess material suitability, considering application requirements and system-level integration.

For nearly two decades, the research laboratories at the University of Zaragoza (UZ) have collaborated with CERN to support the physical characterization of materials under specialized conditions. In particular, the Physical Measurements Service at UZ is equipped with advanced scientific instrumentation dedicated to materials research at low temperatures and under high magnetic fields. The laboratory has recently undergone a major upgrade through two infrastructure projects, funded by the European Union, NextGeneration EU funds EQC2021 call. These projects were specifically focused on enhancing capabilities in the low and very low temperature ranges. The main instruments currently in operation are:

-**PPMS-14T**: A Physical Property Measurement System capable of measuring electrical, thermal, and magnetic properties in the temperature range of 1.9 K to 400 K and in magnetic fields up to 14 T, using a combined NbTi and $Nb_3Sn$ superconducting magnet. It includes a helium-3 refrigerator enabling measurements down to 0.35 K for heat capacity, resistivity, and electrical transport. The system features an open architecture, allowing for the design of custom user experiments.

-**MPMS3 SQUID Magnetometer**: A Superconducting Quantum Interference Device Magnetometer designed to measure magnetic properties (both dc and ac) from 1.8 K to 400 K. It includes a rotator option for anisotropy studies in single crystals, a magneto-optic module for detecting light-induced magnetic effects, and a pressure cell for magnetic measurements under high-pressure conditions. A helium-3 refrigerator extends its temperature range down to 0.4 K for sub-Kelvin magnetic property characterization.

-**Dynacool**: A cryogen-free platform for the measurement of physical properties in magnetic fields up to 9 T. This advanced system enables cutting-edge research in materials science, extending the accessible temperature range into the sub-Kelvin and millikelvin regimes through the use of a helium-3 and dilution refrigerator.

The following sections present the key physical properties of materials and the corresponding testing methods, along with selected examples of studies carried out as part of this long-standing collaboration. These examples are of particular relevance to the mechanical and materials engineering challenges encountered in particle accelerator environments and therefore align closely with the objectives of this course.

## 2   Electrical properties of materials

The classical understanding of electrical conduction is described by Ohm's law, $V = IR$, where resistance $R$ is defined as the ratio of voltage to current. The intrinsic material property associated with resistance is the resistivity, $\rho = RA/L$, which is independent of the specimen's geometry ($L$ is length and A the cross section). Ohm's law applies only to so-called resistive materials, those that exhibit a linear relationship between voltage and current. In reality, the electrical properties of a solid are determined by its electronic band structure. Based on a simplified band model, materials can be categorized into three main classes: metals, insulators, and semiconductors. Metals possess partially filled energy bands, allowing free movement of conduction electrons. Insulators exhibit a wide energy gap between the filled valence band and the empty conduction band, while semiconductors have a narrower band gap, allowing electrons to be thermally excited into the conduction band.

The temperature dependence of resistivity is characteristic of each class of material. In semiconductors, resistivity typically decreases with increasing temperature due to thermal excitation of charge carriers. In contrast, metals show an increase in resistivity with temperature. In metals, this behavior is generally linear and results from enhanced electron scattering due to increased lattice vibrations and defects such as vacancies. In an ideal crystal at absolute zero, where there is no thermal motion and the lattice is perfectly periodic, electrons experience no scattering, and electrical conductivity becomes infinite. However, in real metals, resistivity does not vanish at low temperatures but instead approaches a residual value determined by impurities and structural defects. These factors limit the mean free path of conduction electrons. Therefore, low-temperature resistivity measurements can provide valuable information about the purity and structural quality of a metallic sample.

## 2.1 Testing methods and standards

Testing the resistance of a material is not a difficult task but requires some experimental precautions and procedures to ensure accurate results. Low resistances require the measurement through four-point probe technique, with two leads to supply current and two separate leads to measure voltage, thereby eliminating the contribution of cables resistance and minimizing contact resistance effects. In contrast, for high-resistance values, typically above 10 kΩ, a two-point configuration generally provides sufficient accuracy. Electrical contacts on the sample can be made using various techniques, including soldering, conductive silver paint or silver epoxy, wire bonding or spring-loaded pins, depending on the sample geometry and material properties.

## 2.2 Variation of resistance with temperature, $R$(T). Examples: $T_c$, RRR

A common example of resistance characterization under varying external parameters, such as temperature, magnetic field and current, is the determination of the critical temperature $T_c$ in superconductors. The measured value of $T_c$ is influenced by both the applied magnetic field and the magnitude of the current used during the resistance measurement [1].

The Residual Resistance Ratio (RRR) is defined as the ratio of resistance of a material at room temperature (300K) to its resistance at low temperature, often taken at 10 K, where the resistance is effectively constant and serves as an approximation of the 0 K value. RRR is highly sensitive to the presence of impurities and crystallographic defects, and thus serves as a practical indicator of a material purity and overall quality.

High-purity niobium, with a superconducting transition temperature $T_c$ = 9.2K, is used in the fabrication of high-quality-factor resonant cavities for particle accelerators. The LHC radiofrequency cavities, for example, operate at 4.5 K. Interstitial impurities, such as oxygen, nitrogen, carbon, and hydrogen significantly affect the thermal conductivity of niobium and degrade cavity performance. The total impurity content can be indirectly assessed through RRR measurements. Over the past decade, improvements in melting techniques have led to a substantial increase in the RRR of niobium ingots, from approximately 30 to values exceeding 300.

## 3 Thermal properties of materials

Before delving into details, it is important to highlight the experimental challenges and limitations associated with heat measurements and the physics of heat transfer. Although the measurement of thermal properties is conceptually straightforward, practical implementation is complex due to the difficulty of controlling heat flow and the intrinsic involvement of the sensors within the system under study.

A useful way to appreciate these challenges is to compare thermal conductivity with electrical conductivity, particularly in distinguishing between good and poor conductors or insulators. In the case of thermal conductivity, copper (a commonly used and effective conductor) has a thermal conductivity of approximately 400 W/m·K (silver: 430 W/m·K), while diamond and graphene exhibit much higher values, around 2200 W/m·K and 5000 W/m·K, respectively. On the opposite end, advanced insulators such as silica aerogel blankets have thermal conductivities as low as 0.01 W/m·K. This results in a conductivity range spanning 4 to 5 orders of magnitude. In contrast, electrical conductivity in solids spans a much wider range, from approximately $10^{-20}$ to $10^{7}$ S/m. Excellent electrical conductors like copper and silver exhibit conductivities of $6.0·10^{7}$ and $6.3·10^{7}$ S/m, respectively. Diamond, in contrast, is a good electrical insulator with a conductivity of approximately $10^{-13}$ S/m, and materials such as Teflon can reach values as low as $10^{-20}$ S/m. This results in a range spanning about 27 orders of magnitude. Consequently, electrical resistivity measurements can often be conducted without the measurement setup significantly influencing the result.

Commercial instruments and techniques for measuring thermal properties must address several intrinsic difficulties, including precise control of heat flow, minimization of heat losses via radiation, and accurate detection of small temperature gradients. Achieving reliable, high-accuracy results within practical timeframes requires sophisticated system design and calibration.

In the following sections, the key thermal properties of materials, namely heat capacity and thermal conductivity, are presented. Their physical significance is discussed, along with examples of experimental studies and relevant applications of these measurement techniques.

## 3.1 Heat capacity - HC

The heat capacity of a system is defined as the amount of heat that enters a system divided by the temperature change of the system, $C = \Delta Q/\Delta T$, where $\Delta Q$ is the amount of heat added to the system to raise its temperature by an amount $\Delta T$. It is one of the most fundamental thermodynamic properties, and it allows the derivation of state functions such as enthalpy $H$ and entropy $S$ associated with a process. Heat capacity is related to the internal energy of a system and can be derived from the first law of thermodynamics. It contains the global contribution of all the energy levels of the system, with information about the lattice (phonons) and the electrons, and all the physical processes where those are involved. The heat capacity is typically reported as molar heat capacity or specific heat, $C_P$, where the subscript $P$ indicates that the measurement is performed at constant pressure.

The total heat capacity of a sample can be determined experimentally with high accuracy, however, isolating the individual contributions from different physical mechanisms is more challenging. In solids, the lattice contribution arises from the normal vibrational modes of the crystal structure and can be estimated if the vibrational density of states is known. Debye developed a model that treats the solid as an isotropic, continuous elastic medium, which provides a good approximation for many crystalline materials. Within this model, the Debye temperature $T_D$ corresponds to the cutt-off frequency or maximum frequency for the solid structure and it is a measure of lattice stiffness. For example, soft materials such as lead, have a low Deby temperature ($T_D$ = 105 K), whereas diamond, with its extremely rigid lattice, exhibits a much higher Deby temperature ($T_D$ = 2230 K).

Within the Debye model, the heat capacity exhibits two limiting behaviors at high and low temperatures. At high $T$ ($T >> T_D$), $C_P$ approaches the Dulong Petit limit, given by $C_P = 3NR$, where $N$ is the number of atoms per mole and $R$ = 8.314 J/molK is the gas constant. At low $T$, ($T << T_D$), the heat capacity follows a cubic temperature dependence, $C_P = AT^3$, where the coefficient A is related to the Debye temperature by the expression A = $(12/5)\pi^4 NR/T_D^3$. As a result, in the standard characterization of materials, low-temperature $C_P(T)$ provide a means to determine the Debye temperature, which is directly related to the elastic properties of the material (structural rigidity).

HC measurements are a powerful tool for studying structural and magnetic phase transitions. In materials undergoing such transitions, the heat capacity exhibits an anomaly at the transition temperature, reflecting abrupt changes in enthalpy and entropy associated with the phase change [2]. The total heat capacity is commonly considered as the sum of two components: a continuous background contribution from the degrees of freedom not involved in the ordering process, and an anomalous contribution arising from the degrees of freedom that participate in the transition. In materials that display sharp first-order phase transitions, the latent heat can be determined from heat capacity data. The latent heat corresponds to the enthalpy change associated with the transition and is quantified as the area under the heat capacity anomaly at the transition temperature.

### 3.2 HC testing methods

Various techniques are available for determining the heat capacity of materials, which can be broadly classified into absolute and differential methods. Absolute techniques include adiabatic, semi-adiabatic, and AC calorimetry, all of which involve applying heat to a thermal system that includes the sample, and monitoring the resulting temperature evolution. The heat capacity is then determined by solving the corresponding heat balance equation. Among absolute methods, adiabatic and semi-adiabatic calorimetry commonly employ the heat-pulse technique, a standard and widely used approach. In this method, heat is applied to the system in the form of a square pulse via a heater, and the sample temperature response is analyzed to extract the heat capacity [3]. In the semi-adiabatic method, also known as the relaxation method, the temperature of the sample rises and falls exponentially, following the solution of the differential equation governing the thermal model. The relaxation time, the characteristic time constant of this temperature evolution, is used to determine the heat capacity of the sample. The simplest thermal model can be represented using an electrical circuit analogy, where the heat capacity of the sample is modeled as a capacitor, and the thermal link to the environment, characterized by the thermal conductance of the connecting wires ($K_w$) is modeled as a resistor with resistance $1/K_w$. More advanced models account for additional elements, such as the thermal contact resistance between the sample and the measurement platform or sample holder.

The semi-adiabatic relaxation method offers several advantages over other heat capacity measurement techniques and is the method implemented in the PPMS system at the UZ research laboratories [4,5]. It is faster than adiabatic calorimetry and provides higher accuracy than AC calorimetry. Additionally, it enables measurements on small sample quantities, ranging from as little as 0.1 mg up to 100 mg, in either solid or liquid form. While the method is generally robust, it does have limitations when applied to systems exhibiting sharp divergences in heat capacity, such as those occurring during first-order phase transitions. However, these challenges can be addressed by adapting the measurement protocol and analysis procedures to account for the specific characteristics of such transitions.

### 3.3 Thermal conductivity, *k*

In heat transfer analysis, the thermal conductance of a solid, $K$, is defined as the steady-state ratio of heat flow to the temperature difference across the sample, $K = \dot{Q}_{th}/\Delta T$ W/K. The thermal conductivity for the material $k$ is an intrinsic property related to the conductance by the geometric factors of the sample, $k = K (L/A)$ W/m·K where $L$ is the length and $A$ is the cross section of the sample. Under time-dependent (non-steady-state) conditions, heat transfer is governed by the thermal diffusion equation. In this context, thermal diffusivity characterizes the rate at which heat propagates through a material from the hot end to the cold end, being proportional to the thermal conductivity and inversely proportional to heat capacity (or thermal bulk).

Heat energy can be transmitted through solids via electrical carriers (electrons or holes), lattice waves (phonons), electromagnetic waves, spin waves, or other excitations. In metals electrical carriers carry the majority of the heat. In this sense, thermal conductivity provides valuable insight into the scattering mechanisms affecting electrons and phonons in the material. From the kinetic theory of gases, within a certain approximation, the thermal conductivity can be expressed as: $k = 1/3\ C_P <v><l>$, where $v$ is the average particle velocity and $l$ is the mean free path of a particle between collision.

Normally, the total thermal conductivity can be written as a sum of all the components representing various excitations. In insulators it is determined by their phonon dispersions, the finite excitations of their phonon populations and the lifetimes of these excitations, being acoustic phonons the dominant heat carriers.

Beyond its technological relevance, the measurement of thermal conductivity serves as a powerful tool for investigating fundamental physical phenomena. It enables the study of diverse processes influencing thermal conduction, such as phonon scattering mechanisms, including lattice defects, dislocations, carrier concentrations and interactions, and anharmonic effects, as well as heat transfer at the nanoscale, where thermal conductivity becomes size-dependent. Additionally, the low-energy phonon spectrum can be explored in conjunction with neutron scattering experiments and theoretical calculations.

Similar to heat capacity analysis, many materials can be described within the Debye model approximation, wherein the Debye temperature, $T_D$ establishes the principal energy scale. For insulators, the thermal conductivity typically exhibits a maximum at a temperature, which is a fraction (often one-tenth) of $T_D$. This peak delineates two regimes: above this temperature, phonon-phonon scattering dominates and limits the phonon mean free path, $l_{ph}$, at lower temperatures, collision with defects becomes the predominant limit on $l_{ph}$. The behavior of amorphous solids is quite different due to the absence of translational symmetry. At normal and high $T$, the length of the free path can be considered constant, down to 20-50 K. At lower temperatures, thermal conductivity follows a $T^2$ law, and $k$ deviates from the behavior of the heat capacity $C_P$. This anomaly is attributed to resonant scattering caused by tunneling between energy levels separated by barriers, often modeled as random two-level systems [6].

### 3.4  *k* testing methods

While the measurement of the thermal transport is quite elementary in principle, measurements data are very error prone, time consuming and laborious due to problems in controlling heat flow and accurately measuring small temperature differentials in a convenient manner. In general, there are two basic techniques for measuring thermal conductivity: steady-state methods and transient or non-steady-state methods. Each of these methods is suitable for a limited range of materials, and they are based on the fundamental laws of heat conduction and electrical analogy.

Steady-state methods have been traditionally used due to their mathematical simplicity. There are many techniques available, where a very extended technique is based on a linear heat flow, which is the two-thermometer-one-heater method, where the heater, at one end of the sample generates heat, establishing a steady, linear heat flow through the sample. The temperature gradient is monitored using a hot and a cold thermometer, with the opposite end of the sample connected to a thermal batch acting as reservoir (cold-foot). This technique is commercially implemented in the PPMS as the Thermal Transport Option (TTO), available in the UZ labs [7]. It addresses many experimental challenges by employing optimized sample mounting, small and highly accurate Cernox chip thermometers, and sophisticated software that dynamically models heat flow through the sample while correcting for any heat losses. Thermal conductivity of solid materials is thus measured on small samples via the linear flow method over a temperature range from 2 K to 400 K and magnetic fields up to 14 T. The sample, heater, cold reservoir, and thermometers are housed in an evacuated chamber, minimizing heat losses to primarily radiative mechanisms at the sample periphery. This setup enables measurement of thermal conductivity $k$ and the Seebeck coefficient α (also known as thermopower). Additionally, the system can measure electrical resistivity ρ. These three parameters are critical for determining the thermoelectric figure of merit, ZT = $α^2 T/κρ$, key quantity in thermoelectric materials research.

The selection of sample geometry for the linear heat flow measurement is influenced by the thermal conductivity $k$ of the material. For highly conductive materials with $k$ in the range of 10-150 W/m·K, samples are typically shaped as needles with approximate dimensions of 10 x 1 x 1 $mm^3$. For materials with intermediate thermal conductivity values between 2 and 30 W/m·K, brick-like samples measuring approximately 8 x 2 x 2 mm³ are used. Insulating materials, with *k* ranging from 0.1 to 1.5 W/m·K, are typically prepared as pellets with dimensions around 3 x 5 x 5 mm³. The primary limitations in this technique arise from the maximum heating power available (50 mW) for highly conductive samples and prolonged measurement times for low conductivity samples. The underlying heat balance differential equations are solved within a semi-adiabatic framework to model the time-dependent temperature difference $\Delta T$. In the asymptotic limit, a steady state is achieved where the temperature difference corresponds to the ratio of supplied power $P_0$ to the sample conductance $K_S$, expressed as $\Delta T = P_0/K_S$. The dynamic model further accounts for radiative heat losses from the sample and its connectors, as well as the thermal history of the sample, ensuring accurate determination of thermal conductance.

### 3.5 Examples: NITRONIC 50

An illustrative example of material characterization under extreme conditions was conducted during a test campaign for the ITER project (International Thermonuclear Experimental Reactor). This work was carried out through the long-standing collaboration between the University of Zaragoza and the Engineering Department at CERN, led by A. Arauzo and Stefano Sgobba [8].

The six-storey-tall central magnet of ITER requires a robust support structure to ensure stability and precise alignment under the immense forces generated within the machine. The central solenoid is the strongest of all ITER magnets with a maximum field of 13 T. The supportive "cage" is formed from high-strength, superaustenitic stainless steel (Nitronic 50) components, including tie plates that run the full height of the central solenoid assembly (9 interior and 18 exterior These plates are connected at the top and bottom to upper and lower key blocks, each weighing up to six tonnes.

The material under study was the Nitronic® 50, high strength, high N and Mo austenitic stainless steel used for the ITER Central solenoid compression structure. This steel combines excellent corrosion resistance, low magnetic permeability, and high mechanical strength. These qualities make the material an excellent candidate for many marine environment drive shaft and pump shaft applications and it has been selected for the thermonuclear reactor environment. The tie plates operating at 4 K can be manufactured either from single-piece forgings (SPF), involving blanks exceeding 15 m in length, or as shorter forgings joined by welding (welded solution, WS). Samples from both fabrication routes, obtained from three distinct positions along the tie plates, TOP, MIDDLE, and BOTTOM, were subjected to testing. Physical characterization comprised the measurement of thermal conductivity, heat capacity, electrical resistivity and magnetic properties over the temperature range from 293 K to 4 K. In particular, the thermal properties (see Fig. 1), thermal conductivity and heat capacity, are critical for predicting the quench behavior and cooling dynamics of the large-scale superconducting magnet.

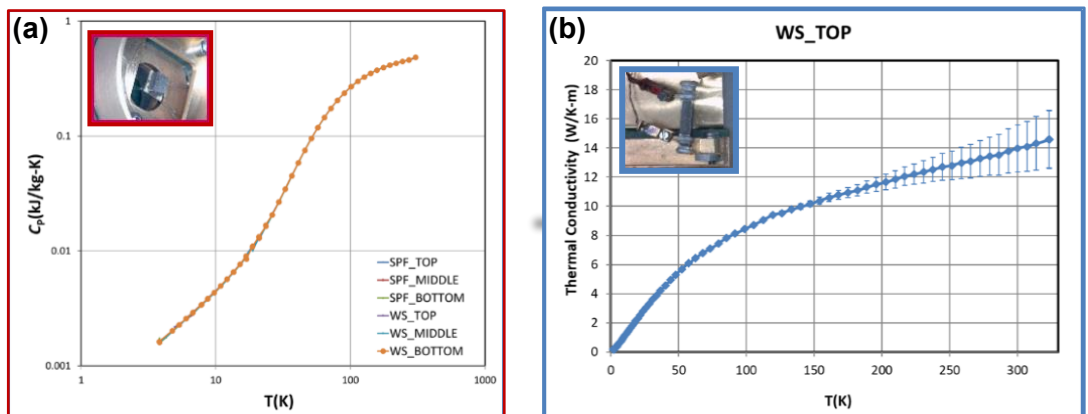

**Fig. 1:** (a) Specific Heat as a function of temperature for all Nitronic50 samples. Logarithmic scale. *Inset*: Photograph of a typical sample mounted in the calorimeter. (b) Thermal Conductivity as a function of temperature. Sample WS_ TOP. *Inset*: photograph of the sample mounted in TTO sample holder. From top to bottom: heater, hot thermometer, cold thermometer, and cold foot.

# 4    Magnetic properties of materials

Magnetism, the phenomenon by which materials assert an attractive or repulsive force or influence on other materials, has been known for thousands of years. However, the underlying mechanisms governing magnetic behavior are complex and were only understood in detail in relatively recent times. Many of our modern technological devices rely on magnetism and magnetic materials; these include electrical power generators and transformers, electric motors, radio, television, telephones, computers, and components of sound and video reproduction systems. While iron, certain steels, and naturally occurring lodestone are commonly recognized magnetic materials, it is less widely known that all substances exhibit some form of magnetic response when exposed to a magnetic field. In fact, every material displays a characteristic magnetic behavior. By studying the variation of magnetization with respect to magnetic field strength, $M(H)$, or temperature $M(T)$, the type of magnetism can be determined. A magnetometer measures the magnetic moment, which quantifies the magnetic response of a sample.

The macroscopic magnetic properties of materials are a consequence of magnetic moments associated with individual electrons. The theoretical foundation of these properties involves quantum-mechanical principles and is inherently complex. All materials exhibit at least one type of magnetic behavior, which can be classified as diamagnetic, paramagnetic, ferromagnetic, antiferromagnetic, ferrimagnetic, superconducting, or superparamagnetic. Diamagnetic and paramagnetic materials are typically considered non-magnetic, as they exhibit magnetization only in the presence of an external magnetic field. In contrast, the other categories result from collective phenomena or long-range magnetic order. The characteristic magnitudes involved in describing magnetic behavior include, magnetic field, $H$[A/m], magnetization, $M$[A/m], $M = \chi \cdot H$, magnetic flux density, $B$[T], $B = \mu H$, magnetic susceptibility, $\chi$, and magnetic permeability, $\mu$, with relative permeability $\mu_r = \mu/\mu_0$. These quantities are related through the expression: $B = \mu_0 H + \mu_0 M = \mu_0(1 + \chi)H = \mu H$ ($\mu = \mu_0(1 + \chi)$).

The magnetic properties of elements and ions are fundamentally determined by their electron configurations. Each electron contributes to magnetism through its intrinsic spin and orbital angular momentum, both of which generate magnetic moments. In atoms with unfilled electron shells, particularly those involving 3d and 4f orbitals, these individual magnetic moments do not cancel, resulting in paramagnetic ions. The magnetic moment of an ion is commonly expressed in units of the Bohr magneton, $\mu_B$, which corresponds to the magnetic moment of a single electron. When such magnetic ions interact in a solid, they may give rise to collective magnetic ordering, forming ferromagnetic, antiferromagnetic, or ferrimagnetic materials. Magnetic order arises from exchange interactions between electron spins, a cooperative effect. Among elemental metals, iron (Fe), cobalt (Co), and nickel (Ni) exhibit ferromagnetic (FM) behavior, the strongest form of magnetism. FM materials are characterized by nonlinear magnetization curves, $M(H)$, exhibiting hysteresis and irreversibility in both field and temperature-dependent magnetization, $M(T)$. Key parameters defining hysteresis behavior include the saturation magnetization, $M_S$, magnetic remanence $M_r$, coercive field $H_c$ and magnetic permeability $\mu$. $M_S$ is an intrinsic property of the material, corresponding to complete alignment of magnetic moments with the applied field. In contrast, remanence and coercivity depend on sample preparation and processing as they are affected by defects and microstructure. Magnetic permeability typically exhibits a maximum as a function of applied field, a feature with significant technological relevance.

Antiferromagnetism arises when neighboring magnetic moments align antiparallel to one another, resulting in a net magnetization close to zero. This leads to a linear magnetization curve $M(H)$ similar to a that of a paramagnetic material, making magnetic susceptibility $\chi$ a useful parameter for characterization. The strength and sign of the exchange interaction responsible for antiferromagnetic (AF) ordering are highly sensitive to both the spacing between magnetic ions and the geometry of the exchange path. Most magnetic insulators exhibit AF rather than FM behavior. A typical AF material consists of two interpenetrating sublattices, each with spins aligned parallel within the sublattice but antiparallel to those in the other, resulting in a cancellation of macroscopic magnetization. Despite this cancellation, the magnetization of each sublattice individually resembles that of a FM. In contrast, ferrimagnetism also involves antiparallel alignment of neighboring magnetic moments; however, the moments differ in magnitude and do not fully cancel, resulting in a net magnetization. As a consequence, ferrimagnets exhibit macroscopic magnetic behavior similar to that of ferromagnets.

A magnetic material is divided into magnetic domains as a minimum-energy configuration. Exchange energy (maintaining parallel atomic magnetic moments) and anisotropy (which establishes the direction of easy magnetization according to each crystal structure) compete against demagnetization energy. Magnetic anisotropy may have different origins, including the intrinsic crystalline structure (magnetocrystalline anisotropy), mechanical deformation coupled with magnetoelastic effects, and by the geometrical shape of the specimen (shape anisotropy). Typically, magnetic domains are microscopic in size. In polycrystalline materials, individual grains may contain multiple domains. However, as grain size is reduced to the nanometric scale, it becomes energetically unfavorable to maintain domain walls, resulting in single-domain behavior. In this regime, nanoparticles can exhibit superparamagnetism, where the entire particle behaves as a single magnetic moment that can fluctuate in response to thermal energy.

## 4.1  Testing methods and standards

Testing the magnetic properties of materials is inherently complex, as numerous experimental conditions can influence the results. To ensure reproducibility and compliance with regulatory requirements, the properties of engineering materials are typically evaluated using standardized methods. These standards vary across national and international standardization bodies.

The testing approach for magnetic materials differs significantly depending on their magnetic properties, particularly their relative magnetic permeability ($\mu_r$). For weakly or **non-magnetic** materials ($\mu_r < 6.0$), a widely used procedure is specified by the American Society for Testing and Materials (ASTM) in standard ASTM A342, Method 5, Standard Test Methods for Permeability of Weakly Magnetic Materials [9]. In contrast, for **strongly magnetic materials**, a closed magnetic circuit is required to perform accurate measurements. In such cases, the standard method recommended by the International Electrotechnical Commission (IEC) is IEC 60404-5 Magnetic Materials - Part 5: Permanent Magnet (Magnetically Hard) Materials - Methods of Measurement of Magnetic Properties.

Commercial SQUID magnetometers and VSM instruments measure magnetic properties under open circuit conditions. Although the magnetic permeability of high-permeability materials cannot be accurately determined in an open circuit, other relevant magnetic parameters, useful for assessing magnet quality, can still be obtained. These measurements can be performed in accordance with IEC 60404-7, Method B, which provides standardized procedures for characterizing magnetic materials in open-circuit configurations.

## 4.2  Examples: non-magnetic alloys, magnetic steels, and manufacturing effects

Magnetic properties of materials intended for use at CERN have been evaluated within the framework of the collaboration above indicated between the UZ and the CERN engineering department. The following examples illustrate representative magnetic tests and corresponding results obtained for both non-magnetic alloys and steels, as well as for magnetic materials, under the relevant testing conditions—primarily cryogenic temperatures and high magnetic fields. These evaluations have included qualification of suppliers and manufacturing processes, with results documented in internal scientific and technical reports submitted by UZ to CERN.

These case studies demonstrate the capabilities of magnetic characterization techniques for both magnetic and non-magnetic samples. They include the determination of intrinsic magnetic properties at low temperatures and high fields, detection of magnetic impurities, assessment of impurity-level variations across suppliers, and evaluation of the influence of processing steps such as cutting, welding, thermal annealing, crystallization, and microstructural changes.

## 4.2.1 NITRONIC50 (ITER project)

Nitronic 50 (N50), the Fe-Cr-Ni alloy used in the tie plate structure of the ITER Central Solenoid coils, exhibits a paramagnetic face-centered cubic (fcc) austenitic phase at room temperature. Upon cooling, a peak in the magnetic susceptibility is observed at approximately 53.5 K, indicating a transition to antiferromagnetic behavior. To meet ITER specifications, the magnetic permeability of this material must remain below µ ≤ 1.03 at cryogenic temperatures

Delta ferrite (bcc FM) may form during solidification of steels and welds, remaining stable at all temperatures. Magnetic measurements confirm that the forged slab material of the welded solution is affected by presence of δ-ferrite (see Fig. 2), confirming previous microstructural observations. In contrast, material from the slab section of the single piece forged plate shows a behavior consistent with a fully austenitic microstructure [8].

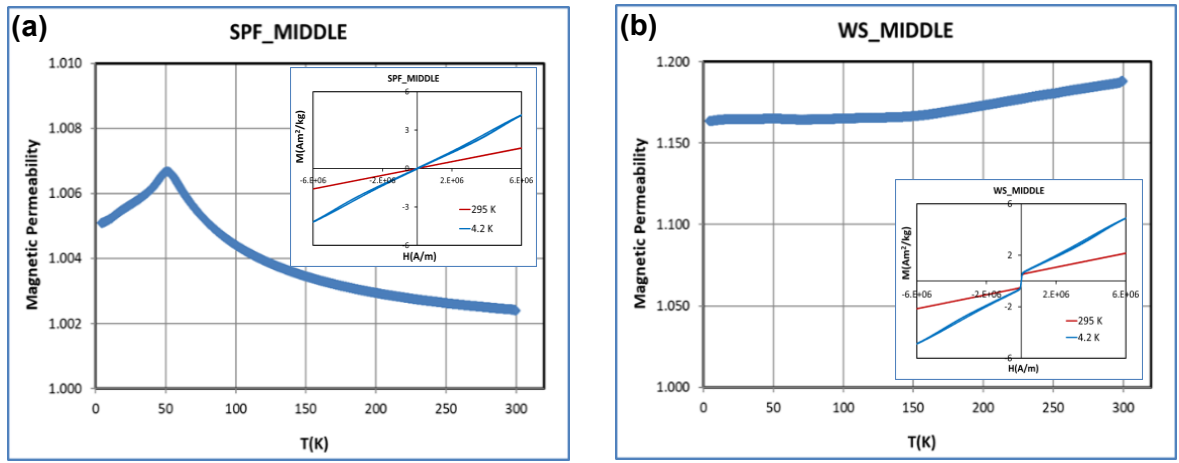

**Fig. 2:** Magnetic permeability as a function of temperature and magnetization hysteresis loops at room temperature and 4.2K (insets) for two specimens of Nitronic50 tie plate at the middle position. (a) Single piece forging, with the behavior expected for a fully austenitic product. (b) Welded solution forging, featuring an increased value of permeability on the whole *T* range, which is a typical effect of presence of δ-ferrite.

## 4.2.2 Non-magnetic tungsten heavy alloy material (ITER project)

Another example of magnetic characterization of non-magnetic materials demanding technical specifications of low magnetic permeability was the evaluation of different suppliers of a tungsten heavy alloy material to be used in the shielding of the ITER fusion reactor. Different suppliers of Inermet® 180, a tungsten heavy alloy (IT180) with a tungsten content > 90% and a NiCu binder phase, were tested. With densities of 17 - 19g/cm$^3$, this material is commonly used for gamma and X-ray shielding or as a collimator, offering high attenuation and reduced shielding thickness due to its density, 60% higher than lead. Low magnetic permeability is essential to avoid interference with the magnetic fields generated in the fusion reactor.

Characterization involved measuring magnetic permeability as a function of temperature and magnetic field, $\mu_r$ (*T, H*), revealing increased levels of magnetic impurities in samples from certain suppliers (see Fig. 3). Continuous quality control of magnetic permeability is essential during the construction phase to ensure compliance with specifications and to monitor production and processing, preventing magnetic contamination.

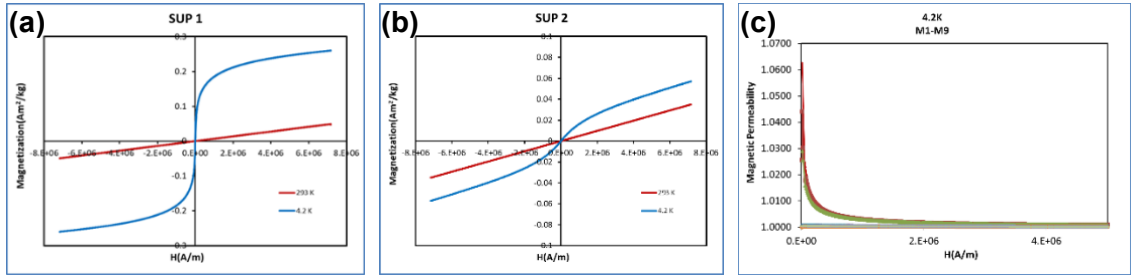

**Fig. 3:** (a), (b) Magnetization hysteresis loops at 293 K and 4.2K for two different suppliers of tungsten alloy. (c) Magnetic permeability at 4.2 K as a function of magnetic field for samples from three different suppliers. The magnetic behavior reveals the presence of magnetic impurities, which are responsible for the elevated magnetic permeability values at low fields. One supplier shows a significantly higher impurity concentration (company names have been omitted for confidentiality).

### 4.2.3 Non-magnetic copper alloys (FCC project)

Different non-magnetic copper alloys, CuAl4Si, CuSn8P and CuSn5 were magnetically characterized as part of the CERN Project FCC Future Circular Collider as potential materials for future 16T dipole magnets. Magnetic permeability was measured down to the equipment's minimum temperature of 1.9 K and up to a maximum magnetic field of $H = 11,141$ kA/m (14 T).

The CuAl7Si alloy showed a magnetic signal attributed to the presence of magnetic impurities that become dominant at low temperature. In contrast, CuSn8P and CuSn5 alloys demonstrated a dominant diamagnetic behavior with an almost negligible magnetic contribution.

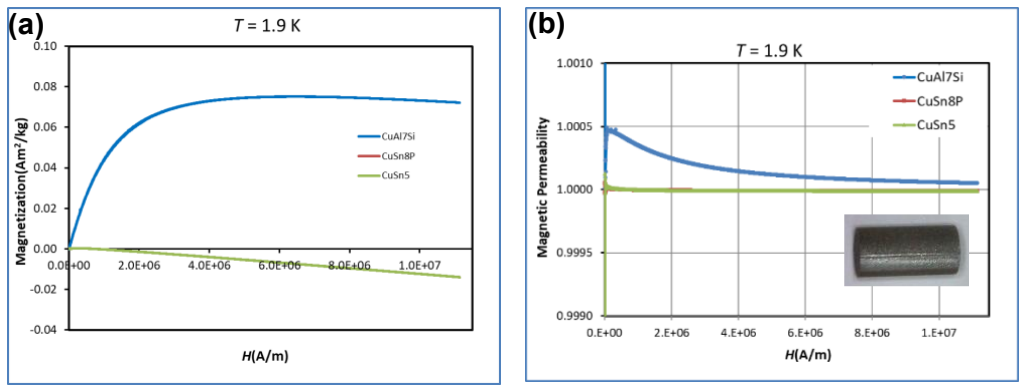

**Fig. 4:** (a) Magnetization as a function of magnetic field at 1.9 K for different Cu alloys exhibiting a diamagnetic contribution (negative slope) in the CuSn8P and CuSn5 samples, while the CuAl7Si sample displays a paramagnetic-like contribution. (b) The corresponding magnetic permeability values quantify the impact of these impurities, with CuAl7Si showing an increased permeability compared to the nearly constant and low values for CuSn-based alloys. The inset shows a typical test sample with a cylindrical geometry: 8 mm in length and 4 mm in diameter.

### 4.2.4 Magnetic steels (FCC project)

Magnetic steels intended for use in the production of magnets for CERN FCC (Future Circular Collider) project have been tested. In a first design phase, these materials are part of the 16 T dipole magnets of the possible future 100 km long accelerator. The materials studied include FeNi36 alloy (INVAR®) and three grades of microalloyed magnetic steels: 300-TG-180, 500-TG-179, and 700-TG-178. These fall under the category of Ferrous Alloys / Magnetic and Electrical Materials. FeNi36 (INVAR®) is a 36% nickel iron alloy known for its extremely low thermal expansion from room temperature up to ~230°C. The microalloyed steels contain trace amounts (0.05-0.15%) of elements such as Nb, V, Ti, Mo, Zr, B, and rare-earth metals to enhance grain refinement and enable precipitation hardening.

At 1.9 K, samples from the three microalloyed steels exhibit very similar magnetic behavior, achieving saturation magnetization values above 200 A·m$^2$/kg and magnetic polarization ($J = \mu_0 M$) around 2.1 T. In contrast, the FeNi36 sample displays lower polarization, consistent with its composition and structure. Magnetic measurements were performed using a VSM (Vibrating Sample Magnetometer) technique in an open magnetic circuit, which limits the accurate determination of absolute magnetic permeability for high-permeability materials due to the demagnetizing factor. In this case, the effective relative permeability is limited to $\mu_r = 8.1$ for all the samples.

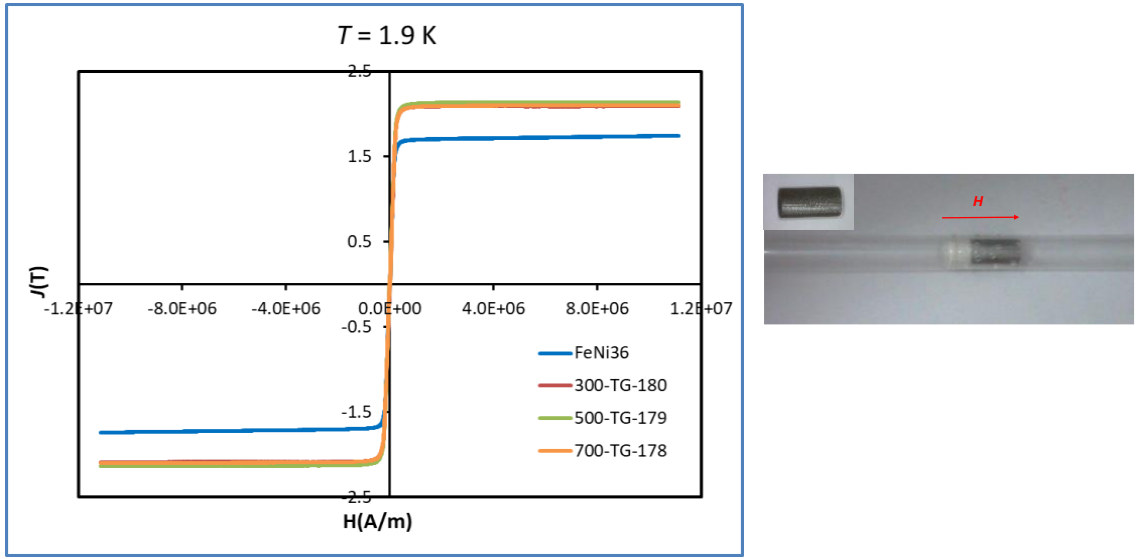

**Fig. 5:** Magnetic polarization loop for the four samples (FeNi36, 300-TG-180, 500-TG-179, and 700-TG-178) measured at $T = 1.9$ K. The photograph displays the sample mounting setup used for magnetic measurements illustrating the cylindrical sample geometry (8 mm length, 4 mm diameter) and the direction of the applied magnetic field.

### *4.2.5  Nanocrystalline soft magnetic materials*

Nanocrystalline soft magnetic materials based on FeSi alloys, such as FINEMET®, show very good properties as soft magnetic materials, following a controlled thermal annealing process. These materials combine the low core loss typically found in amorphous materials with the high permeability and saturation of crystalline materials. The final magnetic performance is achieved through nanocrystallization of the amorphous alloy during the thermal process. The microstructure obtained is made of ultrafine grains (10-20 nm) of Fe and Si within the residual amorphous matrix. The crystallization process starts at about 800 K. The volume fraction of the nanocrystaline phase lies between 50-80%. A second crystallization stage occurs at temperatures exceeding 900 K, during which additional phases may form.

Nanocrystalline Ribbons of FeSiNbCu and Amorphous Ribbons of FeSiB were studied. Magnetization at 1kOe as a function of temperature from 300 K to 1000 K was tested in a special VSM oven to characterize the magnetic properties and the different process taking place. Curie Temperatures are expected in the range 693 K to 833 K. The observed general behavior is displayed in Fig. 6, for one of the samples. In the initial heating cycle, M1(Temp), the first magnetic transition observed can be attributed to the ferro-paramagnetic transition of the **amorphous alloy ($T_c$)**. At higher temperatures it starts an increase of the magnetization in coincidence with the start of the crystallization process and the formation of a magnetic crystalline phase with Curie temperature ($T_{c1}$) superior to the crystallization temperature. This is followed by a drop to zero, marking the Curie temperature of the Fe-Si crystalline phase. A secondary peak near 940–960 K suggests the formation of an additional magnetic phase, likely Fe$_2$B (see Fig. 6 inset).

After the annealing process (thermal cycle up to 1000 K and immediate cooldown), a second measurement, M2(Temp) reveals the Curie Temperature, $T_{c2}$, of the now dominant Fe-Si crystalline phase, along with additional slope changes likely linked to other magnetic phases formed during crystallization.

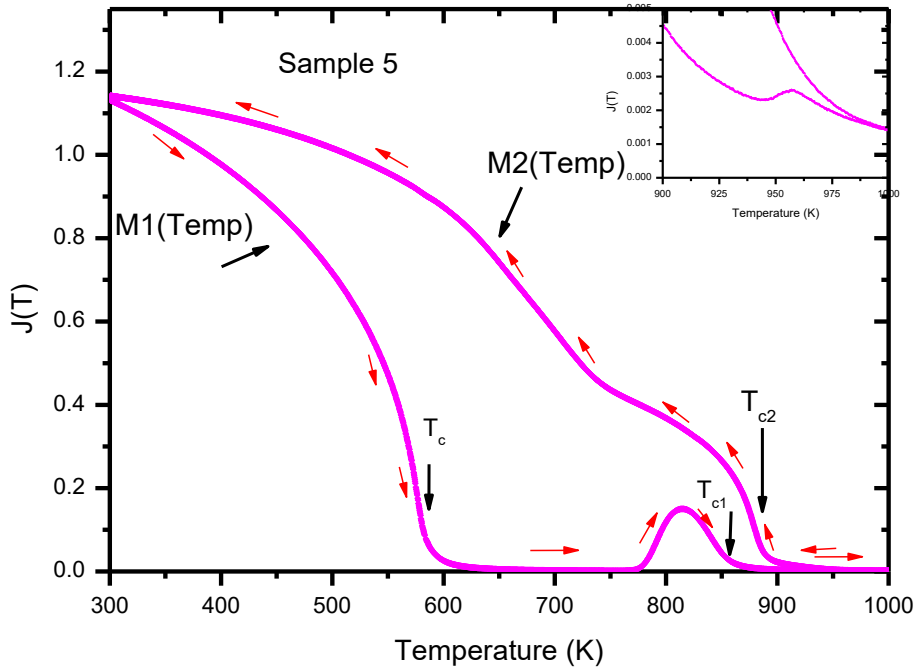

**Fig. 6:** Plot showing typical magnetization curves obtained during heating, M1(Temp) and cooling M2(Temp). Red arrows indicate the measurement process. The main features, $T_c$, $T_{c1}$ and $T_{c2}$ are marked with arrows. *Inset*: small maximum observed at high temperature. The data shown correspond to the magnetic polarization of sample 5, $J(T) = \mu_0 M(T)$.

### *4.2.6 Non-magnetic stainless steels*

Non-magnetic stainless steels are widely used in accelerator applications where magnetic properties must meet strict specifications, particularly regarding maximum magnetic permeability under extreme conditions such as cryogenic temperatures and high magnetic fields. In addition to careful material selection, post-processing steps like cutting and welding play a critical role, as these processes can locally melt the steel and lead to the formation of magnetic phases during rapid solidification.

One example is the magnetic permeability test of P506 austenitic stainless-steel sample with partial copper plating and a longitudinal laser weld along the center. P506 is a nonmagnetic stainless steel, specially developed at CERN belonging to the family of high Mn, high N stainless steels. This special composition allows low relative magnetic permeability ($\mu_r < 1.005$) to be maintained down to cryogenic temperatures. Low-field magnetic measurements reveal an upturn in permeability and small remanence, indicating the presence of magnetic impurities or inclusions. When magnetic permeability is plotted as a function of temperature, a peak is observed at 120 K, in coincidence with the Verwey transition of magnetite ($Fe_3O_4$). This suggests that metal ferrite is formed during the welding process.

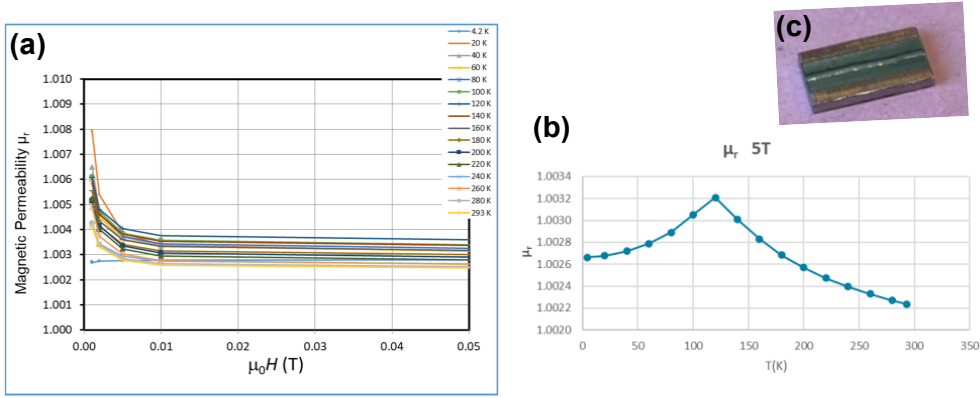

**Fig. 7:** (a) Magnetic permeability as a function of magnetic field up to 5T. The low field range analysis reveals a small magnetic remanence, resulting in increased permeability. (b) Temperature dependence of magnetic permeability from 4.2 K to above 120 K. A distinct peak at 120 K coincides with the Verwey transition of magnetite. (c) Photograph of the laser-welded P506 stainless steel sample used in the measurements.

Another example is the magnetic characterization of AISI 316 (EN 1.4401 – X5CrNiMo17-12-2) non-magnetic stainless steel, which exhibited a noticeable magnetic contribution even at room temperature. This case clearly illustrates how the cutting process can affect magnetic properties. The sample, shaped using EDM (Electrical Discharge Machining), developed a surface layer with magnetic impurities. This affected layer, characterized by a darkened appearance and increased magnetic response, could be removed by surface polishing, restoring the expected non-magnetic behavior (see Fig. 8a). The EDM effect on the surface was also observed by scanning electron microscope (SEM) of a focused ion beam (FIB) cross-section, clearly revealing the affected surface layer on one of the samples (Fig. 8c).

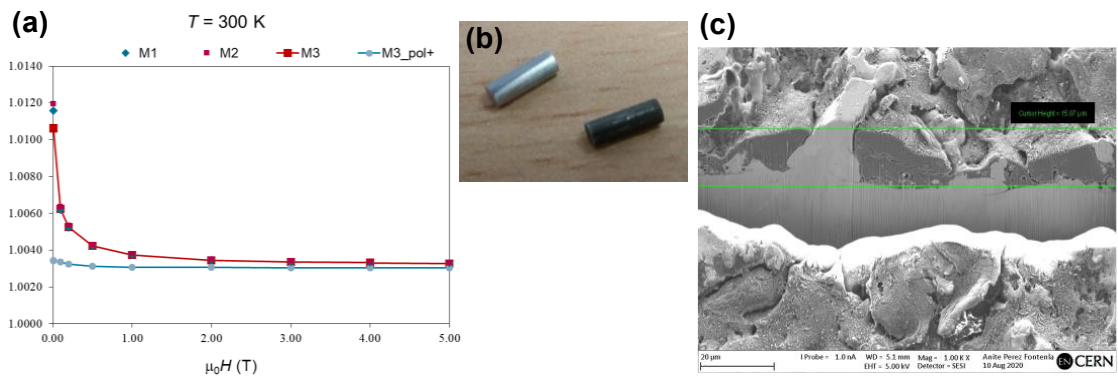

**Fig. 8:** (a) Magnetic permeability at 300 K as a function of applied field up to 5 T for three unpolished samples and one polished sample, illustrating the upturn in permeability caused by EDM cutting on the surface. (b) Photograph showing an unpolished sample (black) alongside a polished sample (shiny metallic). (c) SEM image of a FIB cross-section of the surface of one of the samples revealing the affected layer by the EDM process.

## 5    Conclusions - summary

In summary, material physical properties vary significantly with environmental conditions such as temperature and magnetic field. Therefore, materials intended for particle accelerators operating under extreme conditions, including low temperatures, high magnetic fields, pressure, and radiation, must be rigorously tested. This report highlights various test methods and examples of key physical properties like resistivity, $R(T)$, heat capacity, $C_P(T, H)$, thermal conductivity $k(T)$ and magnetic permeability $\mu_r(T,H)$ based on the capabilities of the UZ physical measurements research laboratory. Multiple case studies demonstrate how characterizing materials at cryogenic temperatures and high magnetic fields is essential to meet technical specifications for final applications. Additionally, the influence of material processing methods, such as annealing, welding, and machining, is illustrated.

Additionally, advanced studies of materials under extreme conditions provide fundamental insights into phase transitions, stability, thermodynamics, defects, and processing, all of which are valuable for materials research.

## Acknowledgement

I wish to thank S. Sgobba and I. Avilés for their collaboration in all the work presented in this report, as well as for their valuable recommendations and comments on the topic.